\documentclass[a4paper,aps,twocolumn,nofootinbib]{revtex4}
\RequirePackage[colorlinks,hyperindex]{hyperref}
\RequirePackage[english]{babel}
\RequirePackage[latin1]{inputenc}
\RequirePackage[T1]{fontenc}
\RequirePackage{mathrsfs}
\RequirePackage{amsmath}
\RequirePackage{amssymb}
\RequirePackage{amsbsy}
\RequirePackage{color}
\RequirePackage{bm}
\hypersetup{colorlinks=true,breaklinks=true,urlcolor=blue,linkcolor=red}
\begin{document}
\title{\bf{Angular-Radial Integrability of Coulomb-like Potentials in Dirac Equations}}
\author{Luca Fabbri$^{\nabla}$, Andre G. Campos$^{\hbar}$}
\affiliation{$^{\nabla}$DIME, Sez. Metodi e Modelli Matematici, Universit\`{a} di Genova, 
Via all'Opera Pia 15, 16145 Genova, ITALY\\
$^{\hbar}$Max Planck Institute for Nuclear Physics, Heidelberg, 69117, GERMANY}
\date{\today}
\begin{abstract}
We consider the Dirac equation, written in polar formalism, in presence of general Coulomb-like potentials, that is potentials arising from the time component of the vector potential and depending only on the radial coordinate, in order to study the conditions of integrability, given as some specific form for the solution: we find that the angular dependence can always be integrated, while the radial dependence is reduced to finding the solution of a Riccati equation so that it is always possible at least in principle. We exhibit the known case of the Coulomb potential and one special generalization as examples to show the versatility of the method.
\end{abstract}
\maketitle
\section{Introduction}
One of the most important problems in physics is finding the solutions in presence of a specific potential for a given equation. Among all possible equations, the Dirac equation is certainly one of the most interesting. And for it, the Coulomb potential is certainly the simplest. Still, the problem of finding exact solutions is always treated in ways that strongly depend on variable separation \cite{CS, CP, GM1, GM2, villalba1994angular, shishkin1989dirac, villalba2002separation, shishkin1989dirac2}, and more in general on various assumptions that tend to limit the range of validity of the method. Meaning, the method used to find solutions with separation of variables for the Coulomb potential can hardly be used to find solutions with no separability of the variables for more general potentials. Still, assumptions like variable separability are very peculiar characters, which do not generally pertain to every situation. Thus, it is necessary to find integrability conditions with weaker assumptions.

In this paper we shall consider this problem. We shall first convert the Dirac equation in the polar form, which is in the form where the spinor field is written in such a way that each component is expressed as a module times a phase. We will consider Coulomb-like potentials, that is potentials arising as temporal components of the vector field and displaying radial dependence. As such they will have the structure of electrostatic spherically symmetric potentials, but we will not limit ourselves to the $1/r$ case solely. Finally, we will pick a specific structure for a trial solution that will enable us to integrate the angular and the radial dependences even when no variable separation will be imposed. In fact, we shall prove that the specific structure of this trial solution recovers variable separability only in the case of the Coulomb potential, and in no other more general case. Then we will present an example of non-separable function that is integrated in both the angular and the radial coordinates simultaneously.

In this respect, it is noteworthy that complete variable separation has been achieved for the Dirac equation in \cite{kruger1991new} at the expense of choosing a spinor representation that is less general than the polar form used here. But in spite of all success that variable separability had granted, this hypothesis generally constitutes a great reduction of the cases in which closed solutions for the radial equations can be found. In the perspective of having exact solutions with a larger degree of applicability, variable separability should be abandoned, and it is therefore necessary to possess methods that even without such a hypothesis could still make one achieve the integrability of the radial equation. In this paper we will present a method in terms of which full integrability can be obtained.
\section{Spinor Fields}
Because we will focus on spinors, it is essential to specify notations and conventions for all quantities involved.

Throughout the paper, we will use the metric $g_{\alpha\nu}$ for the space-time, the tetrads $e^{\sigma}_{i}$ as frame and therefore the Minkowski metric $\eta_{kb}\!=\!g_{\alpha\nu}e^{\alpha}_{k}e^{\nu}_{b}$ for tangent spaces. So, a tensor with Greek indices transforms in terms of some general diffeomorphism (or passive transformations of coordinates) while a tensor with Latin indices transforms in terms of local Lorentz transformations (or active changes of reference systems). One Greek index is raised/lowered in terms of the metric, the passage from Greek to Lorentz indices is achieved in terms of the tetrads and one Latin index is raised/lowered in terms of the Minkowski metric.

In this paper, the Clifford matrices will be $\boldsymbol{\gamma}^{a}$ such that $\left\{\boldsymbol{\gamma}_{a}\!,\!\boldsymbol{\gamma}_{b}\right\}\!=\!2\eta_{ab}\mathbb{I}$ with $\eta_{ab}$ being the Minkowski matrix and $\left[\boldsymbol{\gamma}_{a}\!,\!\boldsymbol{\gamma}_{b}\right]\!=\!4\boldsymbol{\sigma}_{ab}$ will define the infinitesimal generators of a complex Lorentz algebra (in this paper, we specify onto the spin-$1/2$ representation) while $2i\boldsymbol{\sigma}_{ab}\!=\!\varepsilon_{abcd}\boldsymbol{\pi}\boldsymbol{\sigma}^{cd}$ defines the $\boldsymbol{\pi}$ matrix (this matrix is usually denoted as a gamma matrix with an index five, but since in space-time this index has no meaning, and sometimes it may also be misleading, we use a notation with no index). An algebra of Clifford matrices also verifies the following relations
\begin{eqnarray}
&\boldsymbol{\gamma}_{i}\boldsymbol{\gamma}_{j}\boldsymbol{\gamma}_{k}
\!=\!\boldsymbol{\gamma}_{i}\eta_{jk}-\boldsymbol{\gamma}_{j}\eta_{ik}
\!+\!\boldsymbol{\gamma}_{k}\eta_{ij}
\!+\!i\varepsilon_{ijkq}\boldsymbol{\pi}\boldsymbol{\gamma}^{q}
\end{eqnarray}
valid as general geometric identities. By exponentiating the infinitesimal generators $\boldsymbol{\sigma}_{ab}$ for a set of local parameters $\theta_{ij}\!=\!-\theta_{ji}$ we can find the finite local Lorentz group given by $\boldsymbol{S}\!=\!\exp{(-i\alpha\!-\!\frac{1}{2}\theta_{ab}\boldsymbol{\sigma}^{ab})}$ where $\alpha$ is one generic phase. The spinor field $\psi$ is defined as what transforms according to $\psi\!\rightarrow\!\boldsymbol{S}\psi$ in general. With Clifford matrices it is also possible to build a procedure that converts a spinor $\psi$ in its adjoint spinor $\overline{\psi}\!=\!\psi^{\dagger}\boldsymbol{\gamma}^{0}$ so that $\overline{\psi}\!\rightarrow\!\overline{\psi}\boldsymbol{S}^{-1}$ as adjoint transformation. Consequently, with all gamma matrices, and the pair of adjoint spinors, it is possible to construct the bi-linear spinor quantities given by
\begin{eqnarray}
&\Sigma^{ab}\!=\!2\overline{\psi}\boldsymbol{\sigma}^{ab}\boldsymbol{\pi}\psi\\
&M^{ab}\!=\!2i\overline{\psi}\boldsymbol{\sigma}^{ab}\psi
\end{eqnarray}
\begin{eqnarray}
&S^{a}\!=\!\overline{\psi}\boldsymbol{\gamma}^{a}\boldsymbol{\pi}\psi\\
&U^{a}\!=\!\overline{\psi}\boldsymbol{\gamma}^{a}\psi
\end{eqnarray}
\begin{eqnarray}
&\Theta\!=\!i\overline{\psi}\boldsymbol{\pi}\psi\\
&\Phi\!=\!\overline{\psi}\psi
\end{eqnarray}
which are all real tensors. We also have the completeness relationships given by
\begin{eqnarray}
\nonumber
&\psi\overline{\psi}\!\equiv\!\frac{1}{4}\Phi\mathbb{I}
\!+\!\frac{1}{4}U_{a}\boldsymbol{\gamma}^{a}
\!+\!\frac{i}{8}M_{ab}\boldsymbol{\sigma}^{ab}-\\
&-\frac{1}{8}\Sigma_{ab}\boldsymbol{\sigma}^{ab}\boldsymbol{\pi}
\!-\!\frac{1}{4}S_{a}\boldsymbol{\gamma}^{a}\boldsymbol{\pi}
\!-\!\frac{i}{4}\Theta \boldsymbol{\pi}\label{Fierz}
\end{eqnarray}
together with
\begin{eqnarray}
&2\boldsymbol{\sigma}^{\mu\nu}U_{\mu}S_{\nu}\boldsymbol{\pi}\psi\!+\!U^{2}\psi=0\\
&i\Theta S_{\mu}\boldsymbol{\gamma}^{\mu}\psi
\!+\!\Phi S_{\mu}\boldsymbol{\gamma}^{\mu}\boldsymbol{\pi}\psi\!+\!U^{2}\psi=0
\end{eqnarray}
and also
\begin{eqnarray}
&M_{ab}\Phi\!-\!\Sigma_{ab}\Theta\!=\!U^{j}S^{k}\varepsilon_{jkab}\label{A1}\\
&M_{ab}\Theta\!+\!\Sigma_{ab}\Phi\!=\!U_{a}S_{b}\!-\!U_{b}S_{a}\label{A2}
\end{eqnarray}
and
\begin{eqnarray}
&U_{a}U^{a}\!=\!-S_{a}S^{a}\!=\!\Theta^{2}\!+\!\Phi^{2}\label{norm1}\\
&U_{a}S^{a}\!=\!0\label{orthogonal1}
\end{eqnarray}
called Fierz re-arrangement identities.

With the metric $g_{\alpha\nu}$ we define the symmetric connection as
\begin{eqnarray}
&\Lambda^{\sigma}_{\alpha\nu}\!=\!\frac{1}{2}g^{\sigma\rho}(\partial_{\alpha}g_{\rho\nu}
\!+\!\partial_{\nu}g_{\alpha\rho}\!-\!\partial_{\rho}g_{\alpha\nu})
\end{eqnarray}
so that with the tetrads $e^{\sigma}_{i}$ we define the spin connection 
\begin{eqnarray}
&\Omega^{a}_{\phantom{a}b\pi}\!=\!e^{\nu}_{b}e^{a}_{\sigma}(\Lambda^{\sigma}_{\nu\pi}\!-\!e^{\sigma}_{i}\partial_{\pi}e_{\nu}^{i})
\end{eqnarray}
and with gauge potentials $qA_{\mu}$ we can define
\begin{eqnarray}
&\boldsymbol{\Omega}_{\mu}
=\frac{1}{2}\Omega^{ab}_{\phantom{ab}\mu}\boldsymbol{\sigma}_{ab}
\!+\!iqA_{\mu}\boldsymbol{\mathbb{I}}\label{spinorialconnection}
\end{eqnarray}
called spinorial connection. This is needed to write
\begin{eqnarray}
&\boldsymbol{\nabla}_{\mu}\psi\!=\!\partial_{\mu}\psi
\!+\!\boldsymbol{\Omega}_{\mu}\psi\label{spincovder}
\end{eqnarray}
which is the spinor covariant derivative. As well known, the commutator of spinorial covariant derivatives can justify the definitions of space-time and gauge tensors
\begin{eqnarray}
&R^{i}_{\phantom{i}j\mu\nu}\!=\!\partial_{\mu}\Omega^{i}_{\phantom{i}j\nu}
\!-\!\partial_{\nu}\Omega^{i}_{\phantom{i}j\mu}
\!+\!\Omega^{i}_{\phantom{i}k\mu}\Omega^{k}_{\phantom{k}j\nu}
\!-\!\Omega^{i}_{\phantom{i}k\nu}\Omega^{k}_{\phantom{k}j\mu}\\
&F_{\mu\nu}\!=\!\partial_{\mu}A_{\nu}\!-\!\partial_{\nu}A_{\mu}
\end{eqnarray}
which are the Riemann curvature and Maxwell strength.

For the dynamics, we take the spinor field subject to
\begin{eqnarray}
&i\boldsymbol{\gamma}^{\mu}\boldsymbol{\nabla}_{\mu}\psi
\!-\!XW_{\mu}\boldsymbol{\gamma}^{\mu}\boldsymbol{\pi}\psi\!-\!m\psi\!=\!0
\label{Dirac}
\end{eqnarray}
in which $W_{\mu}$ is the axial-vector torsion whereas $X$ is the torsion-spin coupling constant, and this is what is called Dirac equation. If we multiply (\ref{Dirac}) $\boldsymbol{\gamma}^{a}$, $\boldsymbol{\gamma}^{a}\boldsymbol{\pi}$ and then by $\overline{\psi}$ splitting real and imaginary parts gives
\begin{eqnarray}
\nonumber
&i(\overline{\psi}\boldsymbol{\nabla}^{\alpha}\psi
\!-\!\boldsymbol{\nabla}^{\alpha}\overline{\psi}\psi)
\!-\!\nabla_{\mu}M^{\mu\alpha}-\\
&-XW_{\sigma}M_{\mu\nu}\varepsilon^{\mu\nu\sigma\alpha}\!-\!2mU^{\alpha}\!=\!0\label{vr}
\end{eqnarray}
\begin{eqnarray}
\nonumber
&\nabla_{\alpha}\Phi
\!-\!2(\overline{\psi}\boldsymbol{\sigma}_{\mu\alpha}\!\boldsymbol{\nabla}^{\mu}\psi
\!-\!\boldsymbol{\nabla}^{\mu}\overline{\psi}\boldsymbol{\sigma}_{\mu\alpha}\psi)+\\
&+2X\Theta W_{\alpha}\!=\!0\label{vi}
\end{eqnarray}
\begin{eqnarray}
\nonumber
&\nabla_{\nu}\Theta\!-\!
2i(\overline{\psi}\boldsymbol{\sigma}_{\mu\nu}\boldsymbol{\pi}\boldsymbol{\nabla}^{\mu}\psi\!-\!
\boldsymbol{\nabla}^{\mu}\overline{\psi}\boldsymbol{\sigma}_{\mu\nu}\boldsymbol{\pi}\psi)-\\
&-2X\Phi W_{\nu}\!+\!2mS_{\nu}\!=\!0\label{ar}
\end{eqnarray}
\begin{eqnarray}
\nonumber
&(\boldsymbol{\nabla}_{\alpha}\overline{\psi}\boldsymbol{\pi}\psi
\!-\!\overline{\psi}\boldsymbol{\pi}\boldsymbol{\nabla}_{\alpha}\psi)
\!-\!\frac{1}{2}\nabla^{\mu}M^{\rho\sigma}\varepsilon_{\rho\sigma\mu\alpha}+\\
&+2XW^{\mu}M_{\mu\alpha}\!=\!0\label{ai}
\end{eqnarray}
called Gordon-Madelung decomposition identities.

This is the general theory of Dirac spinors presented in a somewhat compact but nevertheless self-contained way as is well known. We will now proceed to translate it in a new form that maintains the same level of generality but which will convert the relevant quantities into other quantities of more immediate interpretation.
\section{Polar Form}
So far we have presented the general theory of spinorial fields, and it is now our goal to transcribe it into the polar form, that is the form in which each component is written as the product of a module times a unitary phase while respecting its Lorentz covariance. To do this it is necessary to clarify that there exist two types of spinors, according to whether they are regular, for which $\Theta$ and $\Phi$ should not be both equal to zero identically, or singular, when $\Theta\!=\!\Phi\!\equiv\!0$ identically \cite{L, Cavalcanti:2014wia, daSilva:2012wp, Ablamowicz:2014rpa, Fabbri:2016msm}. In both cases, the spinorial field equations have been written in polar form, as shown in \cite{Fabbri:2020elt, Fabbri:2016laz, Fabbri:2017pwp}. Singular spinors are important because among them we find Majorana and Weyl spinors, which are fundamental in particle physics. In the present paper we will focus on Dirac fields, that is we will focus on regular spinors. Regular spinors describe general solutions of the Dirac equations for integrable potentials that are fundamental in physics, like the hydrogen atom and the harmonic oscillator \cite{Fabbri:2018crr, Fabbri:2019kyd}. New solutions of the Dirac equations in the free case are presented in \cite{Fabbri:2019kfr}. For a general review we suggest to the reader reference \cite{Fabbri:2020ypd}.

For regular spinors $\Theta$ and $\Phi$ are not identically vanishing at the same time. This makes it possible to see that it is always possible, in the most general case, to write the spinor, in chiral representation, in the form 
\begin{eqnarray}
&\!\psi\!=\!\phi e^{-\frac{i}{2}\beta\boldsymbol{\pi}}
\boldsymbol{S}\left(\!\begin{tabular}{c}
$1$\\
$0$\\
$1$\\
$0$
\end{tabular}\!\right)
\label{spinorch}
\end{eqnarray}
for some complex Lorentz transformation $\boldsymbol{S}$ with $\phi$ and $\beta$ called module and Yvon-Takabayashi angle and where we can appreciate the polar form of each component and the manifest general Lorentz covariance \cite{Fabbri:2016msm}. By considering the polar form of the spinor field, the antisymmetric tensor bi-linear spinor quantities are given according to
\begin{eqnarray}
&\Sigma^{ab}\!=\!2\phi^{2}[\cos{\beta}(u^{a}s^{b}\!-\!u^{b}s^{a})
\!-\!\sin{\beta}u_{j}s_{k}\varepsilon^{jkab}]\\
&M^{ab}\!=\!2\phi^{2}[\cos{\beta}u_{j}s_{k}\varepsilon^{jkab}
\!+\!\sin{\beta}(u^{a}s^{b}\!-\!u^{b}s^{a})]
\end{eqnarray}
showing that they are not independent quantities, as we had already anticipated, because they can always be constructed in terms of the vector bi-linear spinor quantities
\begin{eqnarray}
&S^{a}\!=\!2\phi^{2}s^{a}\\
&U^{a}\!=\!2\phi^{2}u^{a}
\end{eqnarray}
and the scalars bi-linear spinor quantities
\begin{eqnarray}
&\Theta\!=\!2\phi^{2}\sin{\beta}\\
&\Phi\!=\!2\phi^{2}\cos{\beta}
\end{eqnarray}
which show that module and Yvon-Takabayashi angle are a scalar and a pseudo-scalar. Fierz identities reduce to
\begin{eqnarray}
&\!\!\!\!\psi\overline{\psi}\!\equiv\!\frac{1}{2}
\phi^{2}[(u_{a}\boldsymbol{\mathbb{I}}\!+\!s_{a}\boldsymbol{\pi})\boldsymbol{\gamma}^{a}
\!\!+\!e^{-i\beta\boldsymbol{\pi}}(\boldsymbol{\mathbb{I}}
\!-\!2u_{a}s_{b}\boldsymbol{\sigma}^{ab}\boldsymbol{\pi})]
\label{F}
\end{eqnarray}
and all others become trivial except for the ones given by
\begin{eqnarray}
&2\boldsymbol{\sigma}^{\mu\nu}u_{\mu}s_{\nu}\boldsymbol{\pi}\psi\!+\!\psi=0\label{aux1}\\
&is_{\mu}\boldsymbol{\gamma}^{\mu}\psi\sin{\beta}
\!+\!s_{\mu}\boldsymbol{\gamma}^{\mu}\boldsymbol{\pi}\psi\cos{\beta}\!+\!\psi=0\label{aux2}
\end{eqnarray}
and
\begin{eqnarray}
&u_{a}u^{a}\!=\!-s_{a}s^{a}\!=\!1\label{F1}\\
&u_{a}s^{a}\!=\!0\label{F2}
\end{eqnarray}
showing that the normalized velocity vector $u^{a}$ and spin axial-vector $s^{a}$ have only three independent components each, and therefore six in total. The advantage of writing spinor fields in polar form is that the $8$ real components are rearranged into the special configuration in which the $2$ real scalar degrees of freedom, the Yvon-Takabayashi angle and module, remain isolated from the $6$ real components, the spin and velocity, that are always transferable into the frame. The polar form is unique, up to the third axis inversion, which can always be absorbed as generic redefinition of $\boldsymbol{S}$, and up to the discrete transformation $\psi\!\rightarrow\!\boldsymbol{\pi}\psi$, which can always be absorbed as redefinition of the Yvon-Takabayashi angle $\beta\!\rightarrow\!\beta\!+\!\pi$ as clear. It is also necessary to notice that because the spinor is a field, the decomposition to the polar form has to be done in terms of frames that are point-dependent, and hence through a transformation that is local. Therefore, we must expect the parameters of the Lorentz transformation to be local.

As a matter of fact, in general we can write the following expression
\begin{eqnarray}
&\boldsymbol{S}\partial_{\mu}\boldsymbol{S}^{-1}\!=\!i\partial_{\mu}\alpha\mathbb{I}
\!+\!\frac{1}{2}\partial_{\mu}\zeta_{ij}\boldsymbol{\sigma}^{ij}\label{spintrans}
\end{eqnarray}
and consequently
\begin{eqnarray}
&\partial_{\mu}\alpha\!-\!qA_{\mu}\!\equiv\!P_{\mu}\label{P}\\
&\partial_{\mu}\zeta_{ij}\!-\!\Omega_{ij\mu}\!\equiv\!R_{ij\mu}\label{R}
\end{eqnarray}
where $\zeta_{ij}\!=\!-\zeta_{ji}$ are six parameters related to the parameters of the Lorentz transformation. In polar form, the spinor field is reconfigured so that its degrees of freedom are isolated from the components transferable into gauge and frames. When this transfer is done these components are absorbed by gauge potentials and spin connection in such a way that (\ref{P}, \ref{R}) are gauge invariant as well as Lorentz covariant. Therefore $P_{\mu}$ and $R_{ij\mu}$ can be said to be a \emph{gauge-invariant vector momentum} and a \emph{tensorial connection}. With them we can write
\begin{eqnarray}
&\!\!\!\!\!\!\!\!\boldsymbol{\nabla}_{\mu}\psi\!=\!(-\frac{i}{2}\nabla_{\mu}\beta\boldsymbol{\pi}
\!+\!\nabla_{\mu}\ln{\phi}\mathbb{I}
\!-\!iP_{\mu}\mathbb{I}\!-\!\frac{1}{2}R_{ij\mu}\boldsymbol{\sigma}^{ij})\psi
\label{decspinder}
\end{eqnarray}
as the spinorial covariant derivative. We also have
\begin{eqnarray}
&\nabla_{\mu}s_{i}\!=\!R_{ji\mu}s^{j}\label{ds}\\
&\nabla_{\mu}u_{i}\!=\!R_{ji\mu}u^{j}\label{du}
\end{eqnarray}
are general geometric identities. Then taking the commutator of the covariant derivatives we get
\begin{eqnarray}
\!\!\!\!&qF_{\mu\nu}\!=\!-(\nabla_{\mu}P_{\nu}\!-\!\nabla_{\nu}P_{\mu})\label{Maxwell}\\
&\!\!\!\!\!\!\!\!R^{i}_{\phantom{i}j\mu\nu}\!=\!-(\nabla_{\mu}R^{i}_{\phantom{i}j\nu}
\!-\!\!\nabla_{\nu}R^{i}_{\phantom{i}j\mu}
\!\!+\!R^{i}_{\phantom{i}k\mu}R^{k}_{\phantom{k}j\nu}
\!-\!R^{i}_{\phantom{i}k\nu}R^{k}_{\phantom{k}j\mu})\label{Riemann}
\end{eqnarray}
in terms of the Maxwell strength and Riemann curvature, and so they encode electrodynamic and gravitational information, filtering out all information about gauge and frames. We will soon see that it is possible to find $P_{\mu}\!\neq\!0$ and $R_{ij\mu}\!\neq\!0$ such that $F_{\mu\nu}\!=\!0$ and $R_{ij\mu\nu}\!=\!0$ identically in (\ref{Maxwell}, \ref{Riemann}). These solutions encode information related to gauge and frames but with no electrodynamic of gravitational field. Or equivalently, they describe the effect of some action that is not due to any external source.

For the Dirac equations (\ref{Dirac}), plugging into the Gordon decompositions (\ref{vi}, \ref{ar}) the polar form allows us to write the manifestly covariant polar form of Dirac equations
\begin{eqnarray}
&\!\!\!\!\!\!\!\!\!\!B_{\mu}\!\!-\!2P^{\iota}(u_{\iota}s_{\mu}\!\!-\!u_{\mu}s_{\iota})
\!+\!(\nabla\!\beta\!-\!2XW)_{\mu}
\!\!+\!2s_{\mu}m\cos{\beta}\!=\!0\label{dep1}\\
&\!\!\!\!\!\!R_{\mu}\!-\!2P^{\rho}u^{\nu}s^{\alpha}\varepsilon_{\mu\rho\nu\alpha}\!+\!2s_{\mu}m\sin{\beta}
\!+\!\nabla_{\mu}\ln{\phi^{2}}\!=\!0\label{dep2}
\end{eqnarray}
with $R_{\mu a}^{\phantom{\mu a}a}\!=\!R_{\mu}$ and $\frac{1}{2}\varepsilon_{\mu\alpha\nu\iota}R^{\alpha\nu\iota}\!=\!B_{\mu}$ and conversely, from these and (\ref{dep1}, \ref{dep2}) it is possible to obtain the initial Dirac equations, showing that the Dirac equation (\ref{Dirac}) is equivalent to its polar form (\ref{dep1}, \ref{dep2}), as it has been quite extensively discussed in \cite{Fabbri:2016laz}. The spinor equation (\ref{Dirac}) is given by $4$ complex equations, or $8$ real equations, which are as many as the $2$ vectorial equations (\ref{dep1}, \ref{dep2}). Such a pair of vector equations specify all space-time derivatives of both degrees of freedom given Yvon-Takabayashi angle and module. It is very important to say that the discrete transformation given above as $\beta\!\!\rightarrow\!\!\beta+\pi$ requires that also $m\!\rightarrow\!-m$ in order for (\ref{dep1}, \ref{dep2}) to remain invariant.

This concludes the presentation of what is known to be the polar formalism, a way of writing the Dirac spinor field theory that maintains its full generality while allowing for a more immediate interpretation of all relevant quantities and an easier manipulation for practical purposes. One of the practical purposes will be that of finding integrability conditions for exact solutions, as we are going to see later. However, before doing that, we have to present the example of a situation in which $P_{\mu}\!\neq\!0$ and $R_{ij\mu}\!\neq\!0$ but $F_{\mu\nu}\!=\!0$ and $R_{ij\mu\nu}\!=\!0$ in (\ref{Maxwell}, \ref{Riemann}), as promised. We will do that in the following section.
\section{Background Structure}
As promised at the end of the previous section, we now present a non-zero $R_{ijk}$ such that $R^{i}_{\phantom{i}j\mu\nu}\!=\!0$ identically.

We start to specify that we will work in spherical coordinates, with metric given by
\begin{eqnarray}
&g_{tt}\!=\!1\\
&g_{rr}\!=\!-1\\
&g_{\theta\theta}\!=\!-r^{2}\\
&g_{\varphi\varphi}\!=\!-r^{2}|\!\sin{\theta}|^{2}
\end{eqnarray}
and connection
\begin{eqnarray}
&\Lambda^{\theta}_{\theta r}\!=\!\frac{1}{r}\\
&\Lambda^{\varphi}_{\varphi r}\!=\!\frac{1}{r}\\
&\Lambda^{r}_{\theta\theta}\!=\!-r\\
&\Lambda^{r}_{\varphi\varphi}\!=\!-r|\!\sin{\theta}|^{2}\\
&\Lambda^{\varphi}_{\varphi\theta}\!=\!\cot{\theta}\\
&\Lambda^{\theta}_{\varphi\varphi}\!=\!-\cos{\theta}\sin{\theta}
\end{eqnarray}
so that the curvature vanishes.

Nevertheless, a truly simplifying hypothesis is that of exploiting relationships (\ref{ds}, \ref{du}) to find some $R_{ijk}$ which only later be will restricted to be compatible with the $R^{i}_{\phantom{i}j\mu\nu}\!=\!0$ condition. Exploiting (\ref{ds}, \ref{du}) means that a set of assumptions must be taken for the vectors $u_{k}$ and $s_{k}$ in a careful manner. Therefore, to begin, we might demand that $u_{k}$ have temporal and azimuthal components solely, so that due to normalization, they are taken as
\begin{eqnarray}
&u_{t}\!=\!\cosh{\alpha}\label{u1}\\
&u_{\varphi}\!=\!r\sin{\theta}\sinh{\alpha}\label{u2}
\end{eqnarray}
with $\alpha\!=\!\alpha(r,\theta)$ generic function. Orthogonality between vectors $u_{k}$ and $s_{k}$ indicates that we may take $s_{k}$ having the radial and elevational components solely, so that due to normalization, we can take
\begin{eqnarray}
&s_{r}\!=\!\cos{\gamma}\label{s1}\\
&s_{\theta}\!=\!r\sin{\gamma}\label{s2}
\end{eqnarray}
with $\gamma\!=\!\gamma(r,\theta)$ another generic function. Now relations (\ref{ds},\ref{du}) can be solved for $R_{ijk}$ in general. One particular solution is given by
\begin{eqnarray}
&r\sin{\theta}\partial_{\theta}\alpha\!=\!R_{t\varphi\theta}\\
&r\sin{\theta}\partial_{r}\alpha\!=\!R_{t\varphi r}\\
&-r(1\!+\!\partial_{\theta}\gamma)\!=\!R_{r\theta\theta}\\
&r\partial_{r}\gamma\!=\!R_{\theta rr}
\end{eqnarray}
linking the derivatives of the functions $\alpha$ and $\gamma$ to four of the components of the $R_{ijk}$ tensor together with
\begin{eqnarray}
&R_{r\varphi\varphi}\!=\!-r|\!\sin{\theta}|^{2}\\
&R_{\theta\varphi\varphi}\!=\!-r^{2}\cos{\theta}\sin{\theta}
\end{eqnarray}
and
\begin{eqnarray}
&R_{rtt}\!=\!-2\varepsilon\sinh{\alpha}\sin{\gamma}\\
&R_{\varphi rt}\!=\!2\varepsilon r\sin{\theta}\cosh{\alpha}\sin{\gamma}\\
&R_{\theta tt}\!=\!2\varepsilon r\sinh{\alpha}\cos{\gamma}\\
&R_{\varphi\theta t}\!=\!-2\varepsilon r^{2}\sin{\theta}\cosh{\alpha}\cos{\gamma}
\end{eqnarray}
with $\varepsilon$ being a constant after imposing the vanishing of the Riemann curvature.

We also remark that the choice (\ref{u1}, \ref{u2}), (\ref{s1}, \ref{s2}) gives the tetrads
\begin{eqnarray}
&\!\!\!\!e_{0}^{t}\!=\!\cosh{\alpha}\ \ \ \ e_{2}^{t}\!=\!-\sinh{\alpha}\\
&\!\!\!\!e_{1}^{r}\!=\!\sin{\gamma}\ \ \ \ e_{3}^{r}\!=\!-\cos{\gamma}\\
&\!\!\!\!e_{1}^{\theta}\!=\!-\frac{1}{r}\cos{\gamma}\ \ \ \ 
e_{3}^{\theta}\!=\!-\frac{1}{r}\sin{\gamma}\\
&\!\!\!\!e_{0}^{\varphi}\!=\!-\frac{1}{r\sin{\theta}}\sinh{\alpha}\ \ \ \ 
e_{2}^{\varphi}\!=\!\frac{1}{r\sin{\theta}}\cosh{\alpha}
\end{eqnarray}
from which the spin connection is
\begin{eqnarray}
&\Omega_{02r}\!=\!-\partial_{r}\alpha\\
&\Omega_{13r}\!=\!-\partial_{r}(\theta\!+\!\gamma)\\
&\Omega_{02\theta}\!=\!-\partial_{\theta}\alpha\\
&\Omega_{13\theta}\!=\!-\partial_{\theta}(\theta\!+\!\gamma)\\
&\Omega_{01\varphi}\!=\!-\cos{(\theta\!+\!\gamma)}\sinh{\alpha}\\ 
&\Omega_{03\varphi}\!=\!-\sin{(\theta\!+\!\gamma)}\sinh{\alpha}\\
&\Omega_{23\varphi}\!=\!\sin{(\theta\!+\!\gamma)}\cosh{\alpha}\\ 
&\Omega_{12\varphi}\!=\!-\cos{(\theta\!+\!\gamma)}\cosh{\alpha}
\end{eqnarray}
which is obviously a non-trivial spin connection although it gives a zero Riemann curvature tensor.

With the tetrads we can write the tensorial connection in tetradic formalism in the first two indices so that by means of (\ref{R}) and the spin connection we can eventually write that $\partial_{t}\zeta_{12}\!=\!-2\varepsilon$ as easy to see. Consequently, we established the existence of a non-zero tensorial connection for the identically zero Riemann curvature \cite{Fabbri:2018crr}.

The zero Maxwell strength equation $F_{\mu\nu}\!=\!0$ has a much easier form for the non-zero gauge-invariant vector momentum $P_{\mu}$ given by $P_{\mu}\!=\!\nabla_{\mu}\alpha$ in general. But in the present paper we will not make such an assumption, and we will instead pick the general form
\begin{eqnarray}
&P_{t}\!=\!E\!+\!V(r)\label{1}\\
&P_{\varphi}\!=\!-1/2\label{2}
\end{eqnarray}
because we are interested in seeing what happens when a generalized form of Coulomb potential is added.

In general $\varepsilon$ can be non-zero but to have a much cleaner expression for the field equations, we will consider $\varepsilon=0$ from now on and torsion will also be neglected. By using the momentum (\ref{1}, \ref{2}) the field equations (\ref{dep1}, \ref{dep2}) are
\begin{eqnarray}
\nonumber
&\!\!\!\!\!\!\!\!r\partial_{r}\beta\!+\!\partial_{\theta}\alpha-\\
&\!\!\!\!\!\!\!\!-[2(E+V)r\cosh{\alpha}\!+\!\frac{\sinh{\alpha}}{\sin{\theta}}
\!-\!2mr\cos{\beta}]\cos{\gamma}
\!=\!0\label{a}
\end{eqnarray}
and
\begin{eqnarray}
\nonumber
&\!\!\!\!\!\!\!\!\partial_{\theta}\beta-r\partial_{r}\alpha-\\
&\!\!\!\!\!\!\!\!-[2(E+V)r\cosh{\alpha}\!+\!\frac{\sinh{\alpha}}{\sin{\theta}}
\!-\!2mr\cos{\beta}]\sin{\gamma}
\!=\!0\label{b}
\end{eqnarray}
as well as
\begin{eqnarray}
\nonumber
&\!\!\!\!r\partial_{r}\ln{(\phi^{2}r^{2}\sin{\theta})}\!+\!2mr\sin{\beta}\cos{\gamma}
\!+\!\partial_{\theta}\gamma-\\
&\!\!\!\!-[2(E+V)r\sinh{\alpha}\!+\!\frac{\cosh{\alpha}}{\sin{\theta}}]\sin{\gamma}
\!=\!0\label{c}
\end{eqnarray}
and 
\begin{eqnarray}
\nonumber
&\!\!\!\!\partial_{\theta}\ln{(\phi^{2}r^{2}\sin{\theta})}\!+\!2mr\sin{\beta}\sin{\gamma}
\!-\!r\partial_{r}\gamma+\\
&\!\!\!\!+[2(E+V)r\sinh{\alpha}\!+\!\frac{\cosh{\alpha}}{\sin{\theta}}]\cos{\gamma}
\!=\!0\label{d}
\end{eqnarray}
which have to be solved in some special cases.

In the following section we will present one additional assumption for the structure of these equations that will allow us to solve them for a general class of potentials.

In a further section we will consider a specific example of potential in order to illustrate our procedure.
\section{Trial Solution}
It is our task now to consider (\ref{a}-\ref{d}) and see how for general types of Coulomb-like potentials, we may obtain integrability conditions. These four equations are given in terms of the $\alpha$ and $\gamma$ functions, which can be chosen freely. Thus, we will pick the trial solution for which
\begin{eqnarray}
&\sin{\gamma}\!=\!AB\sin{\theta}\label{g1}\\
&\cos{\gamma}\!=\!-A\sqrt{B^{2}\!+\!C^{2}}\cos{\theta}\label{g2}
\end{eqnarray}
and
\begin{eqnarray}
&\sinh{\alpha}\!=\!-AC\sin{\theta}\label{a1}\\
&\cosh{\alpha}\!=\!A\sqrt{B^{2}\!+\!C^{2}}\label{a2}
\end{eqnarray}
as well as
\begin{eqnarray}
&\sin{\beta}\!=\!-AC\cos{\theta}\label{b1}\\
&\cos{\beta}\!=\!AB\label{b2}
\end{eqnarray}
in terms of which $A$ is given by
\begin{eqnarray}
&A\!=\!1/\!\sqrt{B^{2}\!+\!|C\cos{\theta}|^{2}}
\end{eqnarray}
where $B$ and $C$ are generic functions which will have to be obtained as solutions of the field equations. Then
\begin{eqnarray}
\nonumber
&A^{-2}\partial_{\theta}\ln{A}=-B\partial_{\theta}B\!-\!C\partial_{\theta}C|\!\cos{\theta}|^{2}+\\
&+C^{2}\cos{\theta}\sin{\theta}\\
&A^{-2}\partial_{r}\ln{A}=-B\partial_{r}B\!-\!C\partial_{r}C|\!\cos{\theta}|^{2}
\end{eqnarray}
therefore
\begin{eqnarray}
&-\partial_{\theta}\gamma\!=\!\frac{BA^{2}C^{2}}{\sqrt{B^{2}+C^{2}}}
(\partial_{\theta}\ln{X}\cos{\theta}\sin{\theta}+X^{2}+1)\\
&-\partial_{r}\gamma\!=\!\frac{BA^{2}C^{2}\cos{\theta}\sin{\theta}}{\sqrt{B^{2}+C^{2}}}
\partial_{r}\ln{X}\\
&\partial_{\theta}\alpha\!=\!\frac{CA^{2}B^{2}\cos{\theta}}{\sqrt{B^{2}+C^{2}}}
(\tan{\theta}\partial_{\theta}\ln{X}-1-X^{-2})\\
&\partial_{r}\alpha\!=\!\frac{CA^{2}B^{2}\sin{\theta}}{\sqrt{B^{2}+C^{2}}}\partial_{r}\ln{X}
\end{eqnarray}
as well as
\begin{eqnarray}
&\partial_{\theta}\beta\!=\!A^{2}BC\cos{\theta}(\partial_{\theta}\ln{X}+\tan{\theta})\\
&\partial_{r}\beta\!=\!A^{2}BC\cos{\theta}\partial_{r}\ln{X}
\end{eqnarray}
showing that $B/C\!=\!X$ is the only function to be obtained as solution of the field equations. In fact, substituting in (\ref{a}-\ref{d}) gives
\begin{eqnarray}
\nonumber
&-\frac{X^{2}}{X^{2}+1}\tan{\theta}\partial_{\theta}\ln{X}
\!-\!\frac{X}{\sqrt{X^{2}+1}}r\partial_{r}\ln{X}=\\
&=2[(E\!+\!V)r\sqrt{X^{2}\!+\!1}\!-\!1\!-\!mrX]\label{A}
\end{eqnarray}
and
\begin{eqnarray}
\nonumber
&\cot{\theta}\partial_{\theta}\ln{X}\!-\!\frac{X}{\sqrt{X^{2}+1}}r\partial_{r}\ln{X}=\\
&=2[(E\!+\!V)r\sqrt{X^{2}\!+\!1}\!-\!1\!-\!mrX]\label{B}
\end{eqnarray}
and
\begin{eqnarray}
\nonumber
&r\partial_{r}\ln{\phi^{2}}\!+\!2
\!-\![\frac{X}{\sqrt{X^{2}+1}}\partial_{\theta}\ln{X}\cos{\theta}\sin{\theta}-\\
\nonumber
&-2(E\!+\!V)r|\!\sin{\theta}|^{2}X\!+\!2X\sqrt{X^{2}+1}-\\
&-2mr|\!\cos{\theta}|^{2}\sqrt{X^{2}\!+\!1}]A^{2}C^{2}\!=\!0\label{C}
\end{eqnarray}
as well as
\begin{eqnarray}
\nonumber
&\partial_{\theta}\ln{\phi^{2}}+\cot{\theta}
+[\frac{X}{\sqrt{X^{2}+1}}r\partial_{r}\ln{X}+\\
\nonumber
&+2(E\!+\!V)r\sqrt{X^{2}\!+\!1}-\\
&-\frac{X^{2}+1}{|\!\sin{\theta}|^{2}}\!-\!2mrX]A^{2}C^{2}\cos{\theta}\sin{\theta}\!=\!0\label{D}
\end{eqnarray}
as is clear. By combining (\ref{A}) and (\ref{B}) we can see that $X=X(r)$ verifying the field equation
\begin{eqnarray}
&\!\!\!\!\!\!\!\!rX'=-2\sqrt{X^{2}+1}[(E\!+\!V)r\sqrt{X^{2}\!+\!1}\!-\!1\!-\!mrX]\label{ALPHA}
\end{eqnarray}
and when this is substituted in (\ref{C}) and (\ref{D}) we have $\phi^{2}\!=\!Ke^{-G}r^{-2}\sqrt{X^{2}\!+\!|\!\cos{\theta}|^{2}}$ with $G=G(r)$ such that
\begin{eqnarray}
&G'\!=\!2[m\sqrt{X^{2}+1}\!-\!X(E\!+\!V)]\label{BETA}
\end{eqnarray}
with $K$ an integration constant. As a final substitution we can set $X\!=\!-\sinh{(\ln{Z})}$ so that we obtain
\begin{eqnarray}
&Z'=(E\!+\!V\!+\!m)Z^{2}\!-\!(2/r)Z\!+\!(E\!+\!V\!-\!m)\label{Z}\\
&G'\!=\![(E+m)Z^{2}\!+\!Z^{2}V\!-\!(E-m)\!-\!V]/Z\label{G}
\end{eqnarray}
which is treated by solving the first for $Z$ and then plugging the result into the second and performing the integration. Because (\ref{Z}) is a Riccati equation one solution to (\ref{Z}) can be sought after making the transformation
\begin{eqnarray}
Z\!=\!-\frac{z'}{z(E\!+\!V\!+\!m)}
\label{z}
\end{eqnarray}
so that (\ref{Z}) results into
\begin{eqnarray}
&z''\!+\!z'\left(\ln{\left|\frac{r^{2}}{E+V+m}\right|}\right)'\!+\!z[(E\!+\!V)^{2}\!-\!m^{2}]\!=\!0 \label{diff}
\end{eqnarray}
which always admits a solution given as a superposition of two linearly independent solutions. So employing (\ref{z}) one can obtain $Z$ solution of (\ref{Z}) and integrating (\ref{G}) one can find $G$ and eventually the module. Nevertheless, of the two solutions of (\ref{diff}) only one will give a module that will be convergent at infinity. In the following then we are going to focus only on such a convergent solution.

This trial solution has permitted the integration of the angular dependence thus reducing the integration of the radial dependence to finding a solution of (\ref{diff}): because this is a linear differential equation one can always solve it at least in principle. Therefore, this method allows to find solutions for both angular and radial field equations.

We notice that variable separability would be obtained if $X$ did not depend on the radial coordinate, and hence if $Z$ were constant, so that (\ref{Z}) would tell us that
\begin{eqnarray}
&(E\!+\!V\!+\!m)Z^{2}\!-\!(2/r)Z\!+\!(E\!+\!V\!-\!m)\!=\!0
\end{eqnarray}
which implies that the potential must be up to a constant the Coulomb potential. Other potentials would therefore have no variable separability, at least within the range of validity of the trial solution presented here. We are going to comment on this situation more in the conclusions.
\section{An Example}
To give an example of the method we have presented, we consider the Coulomb-like potential
\begin{eqnarray}
&V\!=\!q^{2}/r\!+\!k/r^{2}
\label{pot}
\end{eqnarray}
with $k$ a generic constant. Equation (\ref{diff}) is therefore
\begin{eqnarray}
\nonumber
&z''\!+\!z'\left(\ln{\left|\frac{r^{2}}{E+q^{2}/r+k/r^{2}+m}\right|}\right)'+\\
&+z[(E\!+\!q^{2}/r\!+\!k/r^{2})^{2}\!-\!m^{2}]\!=\!0
\end{eqnarray}
whose solution has to be re-converted by
\begin{eqnarray}
Z\!=\!-\frac{z'}{z(E\!+\!q^{2}/r\!+\!k/r^{2}\!+\!m)}
\end{eqnarray}
and plugged in 
\begin{eqnarray}
\nonumber
&G'\!=\!(E\!+\!q^{2}/r\!+\!k/r^{2}\!+\!m)Z-\\
&-(E\!+\!q^{2}/r\!+\!k/r^{2}\!-\!m)/Z
\end{eqnarray}
to find the $G$ function, so that
\begin{eqnarray}
&\phi^{2}\!=\!Ke^{-G}r^{-2}\sqrt{|\sinh{(\ln{Z})}|^{2}\!+\!|\!\cos{\theta}|^{2}}
\end{eqnarray}
would eventually give the module.

Solutions in general are not easy to find. However, we could study specific limits. In the case where $r$ is large we obtain that the correction becomes negligible compared to the Coulomb potential and henceforth we recover the standard solution for the case of the Hydrogen atom.

Instead, for small $r$ the correction becomes the dominant contribution. In this case
\begin{eqnarray}
z''\!+\!4z'/r\!+\!zk^{2}/r^{4}\!=\!0
\end{eqnarray}
whose solution is given by
\begin{eqnarray}
z\!=\!a\left[\cos{\left(\frac{k}{r}\right)}\!+\!\frac{k}{r}\sin{\left(\frac{k}{r}\right)}\right]
\end{eqnarray}
and consequently
\begin{eqnarray}
Z\!=\!\frac{\frac{k}{r}\cos{\left(\frac{k}{r}\right)}}{\cos{\left(\frac{k}{r}\right)}
\!+\!\frac{k}{r}\sin{\left(\frac{k}{r}\right)}}
\end{eqnarray}
as it is easy to check. Therefore
\begin{eqnarray}
-G\!=\!\ln{r}\!+\!\ln{\left|\ \cos{\left(\frac{k}{r}\right)}\right|
\left|\ \cos{\left(\frac{k}{r}\right)}\!+\!\frac{k}{r}\sin{\left(\frac{k}{r}\right)}\right|}
\end{eqnarray}
and
\begin{eqnarray}
X\!=\!\frac{1}{2}\left[\frac{\cos{\left(\frac{k}{r}\right)}
\!+\!\frac{k}{r}\sin{\left(\frac{k}{r}\right)}}{\frac{k}{r}\cos{\left(\frac{k}{r}\right)}}
\!-\!\frac{\frac{k}{r}\cos{\left(\frac{k}{r}\right)}}{\cos{\left(\frac{k}{r}\right)}
\!+\!\frac{k}{r}\sin{\left(\frac{k}{r}\right)}}\right]
\end{eqnarray}
so that we have
\begin{eqnarray}
\beta\!=\!-\arctan{\left(\frac{\cos{\theta}}{X}\right)}\\
\phi^{2}\!=\!Ke^{-G}r^{-2}\sqrt{X^{2}\!+\!|\!\cos{\theta}|^{2}}
\end{eqnarray}
for the Yvon-Takabayashi angle and module. Finally, it is possible to explicitly write
\begin{eqnarray}
\nonumber
&\phi^{2}\!=\!K\left|\frac{k}{r}\cos{\left(\frac{k}{r}\right)}\right|
\left|\ \cos{\left(\frac{k}{r}\right)}\!+\!\frac{k}{r}\sin{\left(\frac{k}{r}\right)}\right|\cdot\\
&\cdot\sqrt{X^{2}\!+\!|\!\cos{\theta}|^{2}}
\end{eqnarray}
as the expression for the module. We notice that for $r\!\rightarrow\!0$ the module does not have a defined behaviour. This is no problem in itself, as some singularities must be expected for whatever potential that is itself a singular potential.

The potential we used is singular because it was chosen to be the simplest Coulomb-like potential apart from the Coulomb potential itself, with no additional meaning.

Nevertheless, one may study potentials with a positive power of $r$, or more generic functions of $r$, analogously.

Any Coulomb-like potential is integrated by (\ref{g1}-\ref{b2}).

However, we should point out that (\ref{g1}-\ref{b2}) only provide a method to find solutions. They are not an analysis with which to discuss the physical meaning of such solutions.

As a matter of fact it is possible to find many exact solutions of the Dirac equations \cite{Cianci:2015pba,Cianci:2016pvd} which nevertheless do not appear to represent any physical situation.

Our analysis does not provide a discussion for this.
\section{Standard Spinors}
To conclude, it would be instructive to give the explicit expression of the spinor we would have in the standard treatment. This can be done very easily by plugging into the tetrads the explicit expressions of the proposed trial solution given by (\ref{g1}-\ref{a2}) and resulting into
\begin{eqnarray}
&\!\!\!\!\!\!\!\!e_{0}^{t}\!=\!\frac{\sqrt{X^{2}+1}}{\sqrt{X^{2}+(\cos{\theta})^{2}}}\ \ \ \ 
e_{2}^{t}\!=\!\frac{\sin{\theta}}{\sqrt{X^{2}+(\cos{\theta})^{2}}}
\label{e1}\\
&\!\!\!\!\!\!\!\!e_{1}^{r}\!=\!\frac{X\sin{\theta}}{\sqrt{X^{2}+(\cos{\theta})^{2}}}\ \ \ \ 
e_{3}^{r}\!=\!\frac{\sqrt{X^{2}+1}\cos{\theta}}{\sqrt{X^{2}+(\cos{\theta})^{2}}}
\label{e2}\\
&\!\!\!\!\!\!\!\!e_{1}^{\theta}\!=\!
\frac{1}{r}\frac{\sqrt{X^{2}+1}\cos{\theta}}{\sqrt{X^{2}+(\cos{\theta})^{2}}}\ \ \ \ 
e_{3}^{\theta}\!=\!-\frac{1}{r}\frac{X\sin{\theta}}{\sqrt{X^{2}+(\cos{\theta})^{2}}}
\label{e3}\\
&\!\!\!\!\!\!\!\!e_{0}^{\varphi}\!=\!
\frac{1}{r\sin{\theta}}\frac{\sin{\theta}}{\sqrt{X^{2}+(\cos{\theta})^{2}}}\ \ \ \ 
e_{2}^{\varphi}\!=\!\frac{1}{r\sin{\theta}}\frac{\sqrt{X^{2}+1}}{\sqrt{X^{2}+(\cos{\theta})^{2}}}
\label{e4}
\end{eqnarray}
in terms of which we can build the Dirac equation. Such an equation in presence of a Coulomb-like potential $V$ is solved by a spinor field that, in the electron frame of rest and with spin aligned along the third axis, is given in the chiral representation by
\begin{widetext}
\begin{eqnarray}
\psi\!=\!\exp{\left[-i(Et\!-\!1/2\varphi)\right]}
\sqrt{\frac{K\sqrt{X^{2}+|\!\cos{\theta}|^{2}}}{r^{2}e^{G}}}\cdot\exp{\left[\frac{i}{2}\arctan{\left(\frac{\cos{\theta}}{X}\right)}\boldsymbol{\pi}\right]}
\!\left(\!\begin{tabular}{c}
$1$\\
$0$\\
$1$\\
$0$
\end{tabular}\!\right)\label{sol}
\end{eqnarray}
\end{widetext}
for $X$ and $G$ such that
\begin{eqnarray}
&\!\!\!\!rX'=-2\sqrt{X^{2}+1}[(E\!+\!V)r\sqrt{X^{2}\!+\!1}\!-\!1\!-\!mrX]
\end{eqnarray}
and
\begin{eqnarray}
&G'\!=\!2[m\sqrt{X^{2}+1}\!-\!X(E\!+\!V)]
\end{eqnarray}
as it is straightforward to see. These two conditions give the radial integrability, and since the angular integration is already achieved, the equations are fully integrable.
\section{Conclusion}
In this article, we have written the Dirac equation in polar form in presence of Coulomb-like general potentials in spherical coordinates and we have shown that (\ref{g1}-\ref{b2}) result into the integration of the angular dependence and lead to a Riccati equation for the radial dependence: as a consequence, it is always possible at least in principle to find solutions for any form of the Coulomb-like potential.

We noticed that this happens in general, even without assuming variable separation, although variable separability could still be obtained for the Coulomb potential.

It is important to remark that this circumstance seems to contrast with the Peter-Weyl theorem, which states in case of spherically-symmetric potentials that it is always possible to find (linear combinations of) solutions which (at least for each semi-spinor separately) have separability of variables (in this case a spinor can be constructed according to equation (35.1) of \cite{BLP}). In this sense, then our analysis should perfectly fall within the applicability of this theorem and therefore variable separability should have to be recovered. Nevertheless our solution (\ref{sol}) is \emph{not} separable in the sense of equation (35.1) of \cite{BLP}. This apparent contradiction might be resolved if we consider the fact that here our solution is accompanied by a non-trivial tetradic structure (\ref{e1}-\ref{e4}) while the tetrads are trivial in \cite{BLP}. In this case the tetrads would play a very important role in the construction of the Dirac equation.

Opportunity for further investigations might be about finding whether our trial solution is only one of a larger class of solutions for which the full integrability is always ensured, and in this case finding what is the largest class.

More opportunities might be about finding exact solutions for Coulomb-like potential of physical interest, like those that are used for a variety of applications \cite{A1,A2,A3,A4}.

\newpage

\textbf{Data availability statement.} The data that support the findings of this study are available from the corresponding author upon request.

\end{document}